\documentstyle[aps,preprint,epsf]{revtex}

\def\tr{{\rm tr}}

\font\cs=cmcsc10

\begin{document}


\preprint{DUKE-TH-94-73}

\title{Lyapunov Exponent and Plasmon Damping Rate \\
in Nonabelian Gauge  Theories}

\author{T. S. Bir\'o}

\address{Theory Division, MTA KFKI, RMKI, Pf. 49, H-1525 Budapest 114, Hungary}

\author{C. Gong, and  B. M\"uller}

\address{Department of Physics, Duke University, Box 90305,
Durham, NC 27708-0305}

\date{Revised version January 26, 1995}

\maketitle

\begin{abstract}

We explain why the maximal positive Lyapunov exponent of classical SU($N$)
gauge theory coincides with (twice) the damping rate of a plasmon at rest
in the leading order of thermal gauge theory.

\end{abstract}


\section{Introduction}

Numerical studies of Hamiltonian SU($N$) lattice gauge theory in (3+1)
dimensions have shown that the gauge fields exhibit chaotic
behavior in the classical limit  \cite{r1}.  The numerical value of the
largest positive Lyapunov exponent $\lambda_0$ has been obtained for SU(2)
and SU(3) with the result \cite{r1,r2}
\begin{equation}
\lambda_0 = c_N g^2 E_p, \label{1.1}
\end{equation}
where $E_p$ is the average energy per plaquette, $c_2 \approx 0.17$ for
SU(2), and $c_3 \approx 0.10$ for SU(3).  For the SU(2) gauge theory the
complete spectrum of Lyapunov exponents was obtained on small lattices
\cite{r3}.  These calculations, which follow the evolution of a classical
gauge field configuration in Minkowski space, also showed that the
energy density distribution on the lattice rapidly approaches a thermal
distribution \cite{r4}.  This finding confirms the expectation of a finite
growth rate of the coarse-grained entropy density of the gauge field,
which follows from the observation that the sum over all positive
Lyapunov exponents at fixed energy density grows like the volume
\cite{r3}.  Hence, at any given level of coarse-graining, the classical gauge
field ``self-thermalizes'' on a time scale of the order of the inverse
Lyapunov exponent.

In order to determine the value of the maximal Lyapunov exponent $\lambda_0$,
the evolution of the gauge field configurations must be followed over
periods $t_0 \gg \lambda_0^{-1}$. The Lyapunov
exponent is therefore effectively obtained for gauge fields that are
members of a thermal ensemble, and we can identify the average energy
per plaquette $E_p$ in (1) with that of a thermalized lattice.  At
high temperature the gauge field is a collection of weakly coupled harmonic
oscillators, hence the average energy per independent degree of
freedom of the classical gauge field is equal to the temperature $T$,
yielding $E_p = {2\over 3}(N^2-1)T$ for SU($N$).  The factor ${2\over 3}$
accounts for the restrictions imposed by Gauss' law.
We can therefore rewrite the result (\ref{1.1}) as
\begin{equation}
\lambda_0 = {2\over 3} c_N (N^2-1) g^2T \approx \cases{
0.34g^2T &$(N=2)$, \cr 0.53g^2T &$(N=3)$. \cr} \label{1.2}
\end{equation}
As already noted in \cite{r4} these values for $\lambda_0$ coincide, apart
from a factor 2, with those of the damping rate of a thermal plasmon
at rest, obtained by Braaten and Pisarski \cite{r5} in the framework of
thermal perturbation theory:
\begin{equation}
\gamma_0 \approx 6.635 {N\over 24\pi} g^2T = \cases{ 0.176 g^2T
&$(N=2)$, \cr 0.264 g^2T &$(N=3)$. \cr} \label{1.3}
\end{equation}
The goal of the present article is to establish this connection and to
explain the origin of the factor $\lambda_0/\gamma_0=2$.

We approach this goal in several steps. First we review the numerical
``measurement''
of the Lyapunov exponent in classical lattice gauge theory. We point out
that the exponential growth rate of a small perturbation in the magnetic
energy density used in those calculations is equal to twice of the growth rate
of
fluctuations in the elementary field variable, in the continuum limit the
vector
potential. This explains the factor 2 between $\lambda_0$ and $\gamma_0$ .

In the next step we demonstrate that in classical calculations the linear
perturbation
propagation corresponding to the equations of motion of a chaotic dynamical
system
has in general a Fourier spectrum of imaginary frequencies. The Lyapunov
exponent is equal to the magnitude of those imaginary frequencies.

Then we argue that the chaotic dynamics of the classical system acts like
a thermal ensemble averaging the perturbation propagation equation over
stochastic
frequencies. The square of these frequencies can either be positive or
negative. In this case the damping rate and the plasma frequency
of the classical elementary field fluctuations are related to the mean
value and the width of the probability distribution of frequency squares.

The final result of these considerations is that the Lyapunov exponent
as defined in [1] measures
twice the damping rate of classical gauge field fluctuations on the lattice.
It is left to show that the quantum field theoretical calculation of the
thermal damping rate at rest in hot perturbation theory in the leading
${\cal O}(g^2T)$ order survives in the classical $(\hbar \rightarrow 0)$
limit. We begin with the discussion of this point in order to establish
connection with thermal quantum field theory.

\section{Collective Plasma Modes}

We begin by briefly reviewing the derivation of the plasmon damping rate.
Nonabelian gauge field fluctuations in a thermal background have been
studied extensively in the framework of perturbation theory
\cite{r6,r7,r8,r9,r10}.  The gauge field develops massive collective modes
(plasmons) with frequency $\omega(k) >k$ due to interaction with ``hard''
thermal gauge bosons, i.e. excitations with energy of order $T$.  The energy
of a plasmon at rest is $m_g\equiv\omega(0) = {1\over 3}\sqrt{N} gT$
in SU($N$) gauge theory.  For our purpose it is important that the
dispersion relation $\omega (k)$ can be obtained in the framework of
semiclassical transport theory, where classical field fluctuations
$a_{\mu}$ are coupled to the quantized thermal excitations of the
gauge field \cite{r11}.  The gauge invariant description of the collective
modes requires the introduction of effective
$n$-point vertices \cite{r8}, which can be systematically derived from the
effective action \cite{r12,r13}:
\begin{equation}
{\cal L}_{\rm HTL} (a_{\mu})= - {3\over 2} m_g^2 \int d\hat n\; {\tr}
\left(f^{\mu\alpha} {n_{\alpha}n_{\beta}\over (n\cdot D)^2}
f_\mu^{\beta}\right) \label{1.10}
\end{equation}
where $n_{\alpha} = (1, \hat n)$ is a null four-vector, and the integral is
over all directions of the spatial unit vector $\hat n$.  $D^{\nu}$
stands for the gauge-covariant derivative.  We have denoted
the collective gauge potential $a_{\mu}$ and field strength $f_{\mu\nu}$
by lower-case letters to indicate that these describe fluctuations around
a thermal background.  Note that ${\cal L}_{\rm HTL}$ is a classical
construction, with the sole exception that the plasmon rest mass $m_g$
depends on the energy distribution $n(\omega) = (e^{\hbar\omega/T}-1)^{-1}$
of quantized thermal excitations of the gauge field:
\begin{equation}
m_g^2 = {2\over 3} N\; g^2 {\hbar^2\over T} \int {d^3k\over
(2\pi)^3} n(\omega) (1+n(\omega)) = {N\over 9} {g^2\over\hbar} T^2.
\label{1.11}
\end{equation}
At leading order in $g$, (\ref{1.11}) is evaluated for hard thermal
quanta with $\omega = \vert\vec k\vert$.

Braaten and Pisarski\cite{r5} showed that the collective plasmon modes are
unstable due to the effective interaction (\ref{1.10}).
The plasmon damping rate $\gamma(k)$ is defined as imaginary part of
the plasmon pole in the Feynman propagator corresponding to decaying plane
wave solutions. The rate of instability for a plasmon at rest can be
expressed as the imaginary part of the polarization function of the gauge
field at the plasmon pole \cite{r15}:
\begin{equation}
\gamma_0 \equiv \gamma(0) = {1\over 2m_g} {\rm Im}\; {}^* \Pi_{\rm t}
(m_g+i0,0), \label{1.13}
\end{equation}
where the transverse polarization function ${}^*\Pi_{\rm t}(\omega,\vec k)$
only depends on soft modes described by (\ref{1.10}).  The plasmon rest
mass exactly cancels from the expression (\ref{1.13}) and the result
(\ref{1.3}) is a pure number multiplied by $g^2T$, which is a classical
inverse length scale.  In fact, the calculation explicitly makes use of the
classical limit of the Bose distribution, $n(\omega)\to
T/\hbar\omega$, in the evaluation of the loop integral (see eq. (23)
of ref. \cite{r5}).

Since the effective action (\ref{1.10}) can be derived from classical
considerations \cite{r14}, assuming a given spectrum of thermal excitations,
it also applies to the collective excitations of the {\it classical}
gauge field on a lattice.  The sole modification is that the spectrum
of thermal fluctuations is now given by the limit of the
Bose distribution. Denoting the lattice spacing by $a$ we find
\begin{equation}
m_g^2 \to {2\over 3} Ng^2 T \sum_{\vec k} {1\over \omega^2} = {1\over 3\pi}
Ng^2 {T\over a} \label{1.12}
\end{equation}
in the weak-coupling, large volume limit.  The plasmon mass (\ref{1.12})
is a purely classical quantity of dimension (length)$^{-2}$ not
containing $\hbar$, but it diverges in the continuum limit $a\to 0$.
This is not surprising, since the lattice spacing serves as a cut-off
that is required to regularize the ultraviolet divergences of the
classical thermal gauge theory.  The exponential growth rate of small
classical field fluctuations is not affected by this divergence
because it does not depend on the value of $m_g$, as mentioned above.
The result (\ref{1.3}) for the plasmon damping rate $\gamma_0$ remains
valid if the correct plasmon mass $m_g$ in the effective action (\ref{1.10})
is replaced by the value (\ref{1.12}) for the classical gauge field defined
on a lattice.

More intuitively, the independence of $\gamma_0$ from the value of
$m_g$ can be understood as follows.  The cross section for scattering
of a thermal gluon on a slow plasmon is:
\begin{equation}
\sigma \approx {N^2\over N^2-1}\; {g^4\hbar^2\over 4\pi\mu_{\rm D}^2}
\label{1.13a}
\end{equation}
where $\mu_{\rm D}=\sqrt{3} m_g$ is the inverse Debye color screening
length.  The scattering rate $\nu$ is obtained by multiplying with the gluon
density in the initial state and with the Bose factor in the final
state, yielding:
\begin{eqnarray}
\nu &= &2(N^2-1) \int {d^3k\over (2\pi)^3}\,n(\omega) (1+n(\omega))\,\sigma
\nonumber \\
&= &{N^2-1\over N}\; {T\mu_{\rm D}^2\sigma\over g^2\hbar^2} \approx {N\over
4\pi}\; g^2T \approx \gamma_0, \label{1.14a}
\end{eqnarray}
where we have made use of (\ref{1.10}).  From this result, which has
the same structure as the expression (\ref{1.3}) for $\gamma_0$, it is
obvious that the plasmon mass $m_g$ as well as $\hbar$ cancel from the
scattering rate.

\section{Lyapunov Exponents}

The Lyapunov exponents measure the growth rate of infinitesimal
perturbations around an exact solution of the classical lattice
Yang-Mills equations.  Since the maximal Lyapunov exponent $\lambda_0$
was shown to be independent of the lattice spacing, we assume that we
can work in the continuum limit whenever adequate.  If $A_{\mu}(x,t)$
is an exact solution of the Yang-Mills equations, the linearized
equation for a small perturbation $a_{\mu}(x,t)$ around $A_{\mu}$ is
\begin{equation}
D^2 a_{\mu} - D_{\mu} D_{\nu} a^{\nu} - 2i[F_{\mu\nu},a^{\nu}] = 0.
\label{1.4}
\end{equation}
Here $D_{\mu}(A) = \partial_{\mu}-i[A_{\mu}, \quad]$ is the gauge
covariant derivative where the bracket denotes the Lie
algebra commutator, and $F_{\mu\nu}$ is the field strength tensor associated
with the background field $A_{\mu}$.

The numerical approach to the determination of $\lambda_0$ proceeds by
solving (\ref{1.4}) for an arbitrary initial condition $a_{\mu}(x,0)$ and
measuring the growth rate of the norm of $a_{\mu}(x,t)$.  To be precise,
the maximal Lyapunov exponent was determined in \cite{r1,r2} from the
logarithmic growth rate of the ``distance'' between neighboring field
configurations, defined on the lattice as
\begin{equation}
{\cal D}[U'_{\ell},U_{\ell}] = {1\over 2N_p} \sum_p\Big\vert {\tr}\; U_p -
{\tr} \; U'_p \Big\vert, \label{1.5}
\end{equation}
where $U_{\ell}$ are the group valued link variables, $U_p$ denotes the
elementary plaquette operator, and $N_p$ is the total number of
spatial plaquettes.  In the continuum limit, the distance measure
(\ref{1.5}) takes the form
\begin{equation}
{\cal D}[A'_{\mu},A_{\mu}] \propto \int d^3x \left\vert {\tr}\; B'(x)^2
- {\tr}\; B(x)^2\right\vert, \label{1.6}
\end{equation}
where $B(B')$ are the magnetic fields associated with the gauge
potential $A_{\mu}(A'_{\mu})$.  In going from (\ref{1.5}) to (\ref{1.6})
we have suppressed the constant factor $(g^2a/2 N_p)$,
since we are interested only in the growth rate of $(\ln{\cal D})$.  For an
infinitesimal perturbation $a_{\mu}$ that is a solution of the
linearized equation (\ref{1.4}), we obtain:
\begin{eqnarray}
{\cal D}[a_{\mu}\vert A_{\mu}] \equiv{\cal D}[A_{\mu}+a_{\mu},A_{\mu}]
\propto \nonumber \\ \nonumber \\ \int d^3x
\left\vert {\rm tr} \left( {\partial ({\rm tr}\; B^2) \over\partial A_{\mu}}
a_{\mu}\right) + {1\over 2} {\rm tr} \left( {\partial^2 ({\rm tr}\; B^2)\over
\partial A_{\mu}\partial A_{\nu}} a_{\mu} a_{\nu}\right) \right\vert.
\label{1.7}
\end{eqnarray}
The maximal Lyapunov exponent is then defined as
\begin{equation}
\lambda_0 [A_{\mu}] = \lim_{t_0\to\infty} {1\over t_0}
\ln \frac{{\cal D}[a_{\mu}(t_0)\vert A_{\mu}]}{{\cal
D}[a_{\mu}(0)\vert A_{\mu}]}. \label{1.7a}
\end{equation}
In practice, every randomly chosen initial configuration $A_{\mu}(0)$
with a fixed average energy density has been found to yield the same value
for the maximal Lyapunov exponent $\lambda_0$. The numerical calculations
show that the maximal Lyapunov exponent depends only weakly on the lattice
size and extrapolates smoothly to the limit of spatially homogeneous gauge
potentials on a $1^3$ lattice.  We take this as an indication that
$\lambda_0$ is associated with long wavelength perturbations
$a_{\mu}(x,t)$ in an appropriately chosen gauge.

\section{Ergodic Limit}

We now propose to make use of the fact, noted in the Introduction, that
the background gauge field $A_{\mu}(x,t)$ rapidly approaches thermal
configurations, by replacing the {\it long-time} average of the growth
rate of $(\ln {\cal D})$ by the {\it canonical} average over background gauge
fields $A_{\mu}$, where the temperature $T$ is chosen such that the
thermal energy density equals the average energy density of the
time-dependent background field $A_{\mu}(x,t)$.  The replacement of
the temporal average by the canonical average relies on two
conditions: The autocorrelation function of the background field
$A_{\mu}(x,t)$ must decay on a time scale that is short compared with the
time $t_0$ required for the calculation of the Lyapunov exponent, and
the time evolution of the background field must be ergodic on the time
scale $t_0$.

The ergodicity of the background gauge field is assured by its
dynamical chaoticity on time scales long compared to the inverse of
the positive Lyapunov exponents, hence the second condition is fulfilled
\cite{rMa}.  On the other hand, if the first condition were
violated, the Lyapunov exponent would depend on the starting configuration
$A_{\mu}(x,t)$.  In numerical studies [1-4] we have found that this is not
the case. A direct study of the autocorrelation
function performed by us has shown that the first condition is also satisfied.
These conditions are in accordance with the $g^2T \ll gT \ll T$
hierarchy assumed in hot perturbative gauge theory.

The maximal Lyapunov exponent is then obtained from the relation
\begin{equation}
\lambda_0 \approx  {d\over dt}\, \ln\,
\langle {\cal D}[a_{\mu}(t)]\rangle_T,
\label{1.15a}
\end{equation}
where the distance measure (\ref{1.7}) in a thermal background is
\begin{eqnarray}
\langle {\cal D}[a_{\mu}]\rangle_T &\propto& \int d^3x \left\vert {\rm tr}
\left( \left\langle {\partial({\rm tr}\; B^2)\over \partial
A_{\mu}}\right\rangle_T a_{\mu}^{(T)}\right)  \right. \nonumber \\ \nonumber \\
&+& \left. {1\over 2} {\rm tr} \left(
\left\langle {\partial^2 ({\rm tr}\; B^2) \over \partial A_{\mu} \partial
A_{\nu}}\right\rangle_T a_{\mu}^{(T)} a_{\nu}^{(T)}\right)
\right\vert\, . \label{1.8}
\end{eqnarray}
The first term in (\ref{1.8}) vanishes, because
the thermal average of any quantity transforming under the adjoint
representation is zero.  In the second term, the thermal average
projects on to the singlet part of $\partial^2 ({\tr}\; B^2)/\partial
A_{\mu}\partial A_{\nu}$, yielding
\begin{equation}
\langle {\cal D}[a_{\mu}]\rangle_T \propto \int d^3x \left\vert \left\langle
{\partial^2({\rm tr}\; B^2)\over\partial A_{\mu}\partial
A_{\nu}}\right\rangle_T {\rm tr}\; \Bigl(a_{\mu}^{(T)}a_{\nu}^{(T)}\Bigr)
\right\vert.  \label{1.9}
\end{equation}

Since the averaged value of ${\cal D}$ is quadratic in the field fluctuations
$a_{\mu}^{(T)}$ the Lyapunov exponent defined through the magnetic energy
distance measure
is twice as large as the one defined by the dominant exponential growth
rate of the fluctuations of the elementary field
\begin{equation}
\lambda_0 [A_{\mu}] = 2 \,\, \lim_{t_0\to\infty} {1\over t_0}
\ln \frac{||a_{\mu}(t_0)||}{||a_{\mu}(0)||}.
\end{equation}

\section{Classical Spectral Function}

Solving the classical equations of motion one deals with a problem essentially
different from perturbative field theory: instead of investigating transition
amplitudes between scattering states we follow
the evolution of a given initial configuration from a time $t=0$ forwards.
The appropriate method to analyze this evolution is
not the Fourier transformation as in quantum field theory, but the Laplace
transformation. Its inverse transformation is then calculated along a path
which
has all poles of the spectral function on its same side; the path's
position is shifted accordingly, compared with the Fourier transformation.

The classical solution of the equations of motion for field perturbations
therefore
explores in forward time direction all poles of a free oscillator (or wave)
equation. In case of chaotic Hamiltonian dynamics the solutions are both
exponentially growing and damped giving rise to  poles of the
Laplace transform with positive as well as negative real parts.

Making the formal connection between Laplace and Fourier transformation
through a complex rotation of the frequency variable, $s=i\omega$, the inverse
Laplace
transformation path runs {\bf above} all poles in the complex $\omega$-plane.
As a consequence in either case (oscillatory or chaotic) the integration path
for the inverse Laplace
transformation includes {\bf all} poles for positive time and {\bf none}
for negative time while the Fourier transformation includes upper half plane
poles for the advanced (negative time) and lower half plane poles for the
retarded
(positive time) propagator.

\begin{figure}
\vspace{0.15in}
\epsfxsize=6.0cm
\centerline{\epsfbox{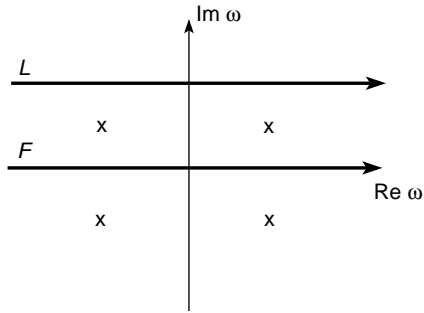}}
\vspace{0.1in}
\caption{The integration paths in the complex frequency plane
  for the inverse Laplace ($L$) and Fourier ($F$) transformations.}
\end{figure}

The position of the poles obtained in a classical time --- forward calculation
may have in general both positive and negative imaginary parts. Therefore
a better quantity for comparison is the spectral function which also
considers poles in the whole complex $\omega$--- plane.

Summarizing this argument the position of all poles of a spectral function
can be obtained from the linearized classical equations of motion for field
perturbations (in the leading order of an $\hbar$ expansion), but
the retarded and advanced propagators used to solve scattering problems in
perturbative field theory discard the unsuitable ones due to their very
definition.
A positive Lyapunov exponent in Hamiltonian (energy conserving) dynamical
systems, on the other hand, always occurs together with its negative
counterpart --- Liouville's theorem ensures it.
Therefore studying positive exponential rates gives an information about
the position of the poles of damped retarded and advanced propagators as well.

The growth or damping rate, or the oscillation frequency of small amplitude
fluctuations in a classical dynamical system is studied by linearizing
the classical equations of motion. This procedure leads to a new differential
operator whose spectrum  gives the poles of the
classical spectral function. Odd parity under time reflection, real valuedness
and
normalization conditions then determine the relative weights of the pole
terms.

The differential operator belonging to the linear perturbation propagation
equation (10)
is identical with the second variation of the classical action, $S''[A]$, taken
at the background field configuration $A$ which is a solution of the classical
equation of motion $S'[A]=0$. Here the prime means variation with respect to
$A$.
Considering the generating functional of connected Green functions, the two
point
function is just the inverse of this differential operator,
\begin{equation}
   G[A,A'] = \langle AA' \rangle - \langle A  \rangle \langle A' \rangle =
(S''[A])^{-1},
\end{equation}
in the Gaussian approximation to the small amplitude fluctuations.
So the linear perturbation propagation in classical equations of motion gives
information
about the saddle point approximated generating functional.

Now aiming at the description of long wavelength plasmon damping we may neglect
spatial derivatives and write the general form of the classical, linearized
perturbation propagation equation (10) schematically as
\begin{equation}
\left[ \frac{d^2}{dt^2} + \Omega^2(t) \right]  a(t) = 0.
\end{equation}
The spectrum of this operator contains two poles on the real axis
$\omega = \pm \Omega$ if $\Omega^2(t)$ is a positive constant. This case,
familiar from zero-temperature perturbative field theory, describes small
oscillations determining the real poles of the spectral function and the
familiar
retarded and advanced propagators.
In classically chaotic, highly excited systems, however, it happens that
$\Omega^2(t)$ is negative. This causes
exponentially growing fluctuations --- a typical source of chaotic behavior.

In order to gain a qualitative understanding about the (classical) spectral
function
of chaotic systems we consider $\Omega^2(t)$ as a Gauss-distributed stochastic
variable \cite{r24}. It can have both negative and positive values, and its
time variation
is replaced by the ensemble variation due to the ergodic property of
classically
chaotic dynamical systems discussed in the previous section.
In this limit the probability distribution of the frequency squares,
$P(\Omega^2)$,
is determined by its two lowest moments,
\begin{eqnarray}
  \langle \Omega^2 \rangle &=& \alpha^2-\gamma^2 \nonumber \\
  \langle \Omega^4 \rangle - \langle \Omega^2 \rangle^2 &=& 4\alpha^2 \gamma^2
\end{eqnarray}
parametrized by two real parameters $\alpha$ and $\gamma$. This parametrization
reflects the fact
that while $\langle\Omega^2\rangle$ can either be positive or negative, the
width of its
distribution is always positive.

The stochastic average of the differential operator for the fluctuations
has to be carried out on the quadratic level, because with the Gaussian
distribution
we assumed white noise property of the stochastic quantity.
We get
\begin{eqnarray}
  \left\langle \left( \omega^2 - \Omega^2 \right)^2 \right\rangle =
  \omega^4 - 2 \langle \Omega^2 \rangle \omega^2 +  \langle \Omega^4 \rangle =
\nonumber \\
   (\omega - \alpha - i\gamma)(\omega - \alpha + i\gamma)
   (\omega + \alpha - i\gamma)(\omega + \alpha + i\gamma).
\end{eqnarray}
This result exhibits the symmetric four pole structure typical for a spectral
function describing classical plasma oscillations
\begin{eqnarray}
  {\cal A}(\omega)  = \frac{1}{4i\pi\omega} && \left(
  \frac{1}{\omega - \alpha - i\gamma} - \frac{1}{\omega - \alpha + i\gamma}
\right.
 \nonumber \\
  &+& \left.
  \frac{1}{\omega + \alpha + i\gamma} - \frac{1}{\omega + \alpha - i\gamma}
  \right)
\end{eqnarray}
yielding the Lorentz shape
\begin{equation}
 {\cal A}(\omega) = \frac{1}{\pi}
\frac{2\gamma\omega}{(\omega^2-\alpha^2-\gamma^2)^2 +
   4\gamma^2\omega^2}.
\end{equation}
The relative signs of the pole terms follow from the definition of the spectral
function
as the difference between the advanced and retarded propagators and from its
odd
time parity ${\cal A}(-\omega) = - {\cal A}(\omega).$ The normalization factor
$1/2\omega$ ensures that
\begin{equation}
\int_{-\infty}^{\infty} \limits \! d\omega \,\, \omega {\cal A}(\omega) = 1,
\end{equation}
so in each mode exactly one boson is counted by the spectral function ${\cal
A}(\omega)$
\cite{r22}.

This particular, four pole spectral function describes a general solution of
the
stochastically averaged perturbation propagation equation which behaves like
\begin{equation}
 a(t) = Ae^{i\alpha t - \gamma t} + Be^{-i\alpha t- \gamma t} +
 Ce^{i\alpha t + \gamma t} + De^{-i\alpha t + \gamma t}.
\end{equation}
After some initial oscillations the exponential growth dominates
the long time  behavior of $|a(t)|$. It is exactly this, which has been
seen in numerical calculations. The conclusion of this argument is that
the Lyapunov exponent of elementary field fluctuations averaged ergodically
is equal to the classical gluon damping rate as expressed by the imaginary part
of the pole positions in the spectral function ${\cal A}(\omega)$.

We note that in a recent publication \cite{r23} a similar Gaussian model for
the chaotic instability in general Hamiltonian flows has been investigated.
Our result presented above recovers the more general one of ref \cite{r23}
for vanishing expectation value of the noisy oscillator frequency square
($\alpha=\gamma)$ after substituting a characteristic timescale $\tau=1/\gamma$
in the general formula (19) of ref. \cite{r23}.

\section{Classical and Quantum Gluon Damping}

Finally we argue again that the leading order gluon damping rate (3) obtained
in hot
perturbative QCD (pQCD) is classical, i.e. it retains its value in the
classical
limit $\hbar \rightarrow 0.$ This fact has been argued before in section II.
Here we briefly reconstruct the argument and resolve some technical issues.
This concludes our reasoning about the equality of
the Lyapunov exponent of chaotic classical lattice gauge theory and the gluon
damping rate at rest in a hot plasma.

The gluon damping rate in hot pQCD is obtained from the definition
(6) dividing the imaginary part of the self energy by the thermal gluon mass
$m_g=gT/\sqrt{\hbar}.$ The general one-loop form of the self energy
contains an integral over hard momenta, a factor of $g^2$ and the phase space
distribution of thermal gluons
\begin{equation}
{\rm Im} \Pi(m_g,0) = g^2 \hbar \int \! d^4k \, \frac{1}{k^2} n(k) f(k/m_g),
\end{equation}
where the  complicated algebraic expression $f(k/m_g)$ depends only on scaled
momentum variables. Using now the long wavelength approximation the phase space
distribution of thermal gluons is replaced with its classical counterpart,
$n(k) \approx T/\hbar\omega,$ leading to
\begin{equation}
{\rm Im} \Pi (m_g,0) \propto g^2T \int d\omega f(\omega/m_g).
\end{equation}
Scaling the integration variable with the Debye mass --- which is of quantum
origin containing the Planck constant --- we see that the imaginary part of
the 1-loop gluon self energy in a hot plasma is proportional to $m_g$.
It follows that the gluon damping rate obtained using ``classical''
thermal gluons does not depend on the Debye mass and Planck's constant,
\begin{equation}
\gamma = g^2T \int dx f(x),
\end{equation}
showing that the result (3) is essentially classical.

Finally, it is still to show whether non-pole contributions to the self energy
in the field theoretical calculation do not interfere with the above arguments.
The one-loop spectral function used there as an input contains a pole term
picking up the zeroes $\omega(k)$ of the inverse propagator corresponding
to collective plasma modes to the lowest order and a cut term describing the
effect of
scattering on thermally excited spacelike modes:
\begin{equation}
{\cal A}(k,\omega)=Z(k) \delta(\omega^2-\omega(k)^2) \, + \,
\beta(k,\omega) \Theta(k^2-\omega^2).
\label{SPECTRAL}
\end{equation}
The cut coefficient $\beta(k,\omega)$ is related to the real and imaginary
parts of the self energy $\Pi (k,\omega)$:
\begin{equation}
\beta(k,\omega) = \frac{{1\over\pi} {\rm Im}\,\Pi}
{\left( k^2-\omega^2+{\rm Re}\,\Pi \right)^2
+ \left( {\rm Im}\,\Pi \right)^2 }.
\label{BETA}
\end{equation}
The respective self energies for the transverse and longitudinal excitations
to leading order in hot perturbative QCD are\cite{r17}
\begin{equation}
\Pi_{\rm t}(k,\omega) = m^2 x^2 \left(1+\frac{1-x^2}{2x}{\rm ln}\frac{1+x}{1-x}
\right) + {i\pi\over 2}m^2 x (1-x^2)
\label{PiT}
\end{equation}
and
\begin{equation}
\Pi_{\ell}(k,\omega) = k^2\, + \, m^2 \left( 2-x\,\ln\frac{1+x}{1-x} \right)
 - i \pi m^2 x  \label{PiL}
\end{equation}
with $x=\omega/k$ and $m^2=3m_g^2/2$. Using these forms one obtains the
following cut parts of the retarded Fourier transform of the spectral function,
$\Delta(t,k),$
for small $k/m$
\begin{eqnarray}
\Delta_{\rm cut,t}(k,t) \rightarrow -2 \Theta(t) \frac{k}{m^2}
\nonumber \\
\int_0^1 dx \, \frac{4x(1-x^2) \sin ktx}{4x^4\left(1+\frac{1-x^2}{2x} \ln
\frac{1+x}{1-x} \right)^2 \, + \, x^2(1-x^2)^2 }
\label{DeltaT}
\end{eqnarray}
and
\begin{equation}
\Delta_{{\rm cut},\ell}(k,t) \rightarrow -2 \Theta(t) \frac{k}{2m^2}
\int_0^1 dx \, \frac{x \sin ktx}{\left(2-x\, {\rm ln}
\frac{1+x}{1-x} \right)^2 \, + \, x^2 }\, .
\label{DeltaL}
\end{equation}
Since the integrand is bounded, the cut contribution
cannot grow exponentially with time and hence does not
contribute to the maximal Lyapunov exponent (\ref{1.15a}). In fact,
the cut contribution vanishes in the long wavelength limit $k\to 0$.

This leaves us with the pole part, which remains finite in this limit.

\section{Summary}

This concludes our argument establishing a connection between
the classical Lyapunov exponent and the gluon damping rate in
hot perturbative QCD.
We note that some elements of the argument are heuristic,
in particular, the replacement of the long-time average of the growth
rate of fluctuations around a specific field configuration by the thermal
average.  This reasoning assumes that the growth rate, or equivalently
the plasmon damping rate, depends only on coarse-grained properties of
the gauge field.  We believe that this is so, because the one-loop
calculation of the damping rate $\gamma_0$ only involves soft loop momenta
\cite{r8} and hence does not depend on details of the short-distance
fluctuations of the gauge field.

Because of the general nature of our argument, we conjecture
that the complete spectrum of Lyapunov exponents obtained in \cite{r3}
reflects the spectrum of damping rates $\gamma(k)$ of excitations in a
thermal bath.  If this were true, it would confirm our assumption that
$\gamma(k) \le \gamma_0$.  Since, at present, it is not known whether
$\gamma(k)$ is a quantity with a classical limit for $k\not= 0$, the
identification with the Lyapunov spectrum remains a conjecture.  We
finally note that if the correspondence between ergodic and canonical
averages holds up for other physical quantities, transport coefficients
of nonabelian gauge fields at the classical scale $(g^2T)$, such as magnetic
screening \cite{r18} or color diffusion \cite{r19}, could possibly also be
calculated by real-time evolution of classical gauge fields on a lattice.
\bigskip

\noindent {\bf Acknowledgements:}
We thank U. Heinz, S. G. Matinyan, H. B.
Nielsen, G. K. Savvidy, and M. Thoma for illuminating discussions
and the referee for pointing out a mistake in our original manuscript.
This work was supported in part by the U.S. Department of Energy
(grant {\cs de-fg05-90er40592}) and in part by the Collaboration
Agreement between the Norwegian Research Council (NFR) and the
Hungarian Academy of Science (MTA) (grant {\cs 422.92/001}).
One of us (B.M.) thanks
the Physics Department of Tokyo Metropolitan University, especially
H. Minakata, for their hospitality and support during his visit there.
T.S.B. acknowledges the hospitality and the support of Physics
Department of Bergen University and of L. P. Csernai during his visit
there.


\begin{thebibliography}{lbl}
\bibitem{r15} The sign conventions in ref. 5 are such that a positive
imaginary part of ${}^*\Pi$ implies a negative imaginary part of the pole
energy.  In writing (6) we have already continued ${}^*\Pi_{\rm t}$ to
Minkowski space.
\end{thebibliography}
\end{document}